\documentclass[11pt,twoside]{article}

\usepackage{asp2004}
\usepackage{epsf}
\usepackage{psfig}
\usepackage{lscape}
\usepackage{graphicx} 

\begin{document}
\title{Short term variability in Be stars and binaries in the Magellanic Clouds.}   
\author{C. Martayan, M. Floquet, A.-M. Hubert, M. Mekkas}   
\affil{GEPI/UMR8111-Observatoire de Meudon, 5 place Jules Janssen 92195 Meudon
cedex FRANCE}    

\begin{abstract} 
We observed a large sample of B \& Be stars in the Magellanic Clouds, 
respectively 176 and 344 stars, with the  multi-object spectrograph GIRAFFE at
ESO/VLT.  We obtained spectra at different settings at various epochs.  It
allowed us to discover several spectroscopic binaries with short  and
intermediate periods. By cross-correlation in coordinates with the  MACHO
database, we found the light-curves for 350 stars in our samples.  Among them,
19 (14 new) are photometric binaries (5 in the LMC, 14 in the SMC).  Among
these 19 photometric binaries 6 are also spectroscopic binaries (5 in the LMC, 
1 in the SMC) and 2 others are Be stars (2 in the SMC). For all these binaries 
the orbital period was determined. Among our sample, 134 Be stars were
observed  by MACHO and, for these stars, we searched for short-term photometric
variability.  We have found 13 objects among Be stars in the SMC which
present short-term  photometric variability (P$<$2.5d) with amplitude lower than
0.1 magnitude.  This short-term variability is often superimposed to a
long-term variation. 
\end{abstract}



\section{Observations}
Thanks to the VLT-FLAMES/GIRAFFE facilities in MEDUSA mode, a large sample of 
B and Be stars have been observed in the LMC-NGC2004 and SMC-NGC330 regions 
and their surroundings, 176 and 346 hot stars respectively. We used a medium 
resolution R=6400 in the blue wavelength range around H$\gamma$ (LR02) and
R=8600 in the  red one around H$\alpha$ (LR06). Among this sample 178 Be stars
were observed.   We have cross-correlated the coordinates of observed stars with the
MACHO and OGLE  databases and we found light curves for 350 stars. Among them
134 are Be stars.\\

Blue spectra of B and Be stars in the LMC were obtained   4 days before red
ones and it was therefore possible to study  the variation of the stellar
radial velocities and to detect  spectroscopic binaries. Despite the small
number of spectra for each  object it was possible to estimate the mass ratio
and the systemic velocity  for some binaries (see Martayan et al. 2005a). By
cross-correlation with  the MACHO and OGLE databases we detected 19 photometric
binaries.  Among them 14 are new photometric binaries (5 in the LMC, 9 in the
SMC).  Among these 19 photometric binaries 6 are also spectroscopic binaries 
(5 in the LMC, 1 in the SMC) and 2 others are Be stars (2 in the SMC).  For
eclipsing binaries, the orbital period was determined with an  accuracy
of 0.001d.

\section{Binaries}
\subsection{Binaries in the LMC}

\begin{figure*}[ht]
\centering
\includegraphics[width=4.5cm, height=9cm,angle=-90]{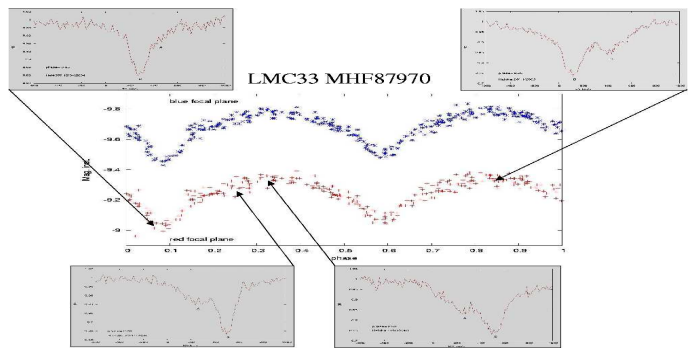}
\caption{Binary LMC33-MHF87970, P=7.117d, MACHO60.7950.16. The central panel
shows the MACHO lightcurves in phase. The four little panels show the GIRAFFE
spectra and their correspondence with the phases.}
\label{binMHF87970}
\end{figure*}

Thanks to observations with VLT-GIRAFFE spectrograph,  we find 23 new binary
systems among the B stars of our sample  (see Martayan et al. 2005a). The
orbital periods were found thanks  to the MACHO data. Results concerning
eclipsing binaries are given  in Table~\ref{binaires}. As an example, the case of the LMC
spectroscopic binary (SB2)  MHF87970 (Porb=7.117d) is illustrated in
Figure~\ref{binMHF87970}. Spectra obtained  with the VLT-GIRAFFE are located at their
orbital phases.  The other eclipsing binaries are shown in Martayan et al.
(2005a)

\subsection{Binaries in the SMC}
Contrary to the LMC, the observations in the blue and red  wavelength have been
done the same night and we cannot detect the variation of radial velocity. We
find 14 eclipsing binaries. Periods were determined and compared to previous
studies. Results are given in Table~\ref{binaires}. The
nature of SMC5\_49816 is reconsidered and we propose SMC5\_49816  like a WUMa
binary because of its very short orbital period.

\begin{table*}[tbph]
\centering
\scriptsize{
\caption{Characteristics of eclipsing binaries in the Magellanic Clouds.
The first column gives the name of the star, the second column gives the 
period of eclipses with an accuracy of$\pm$0.001d . The third column gives the
ratio of masses and the last column gives different comments about the binary.}
\begin{tabular}{@{\ }c@{\ \ \ }c@{\ \ \ }c@{\ \ \ }c@{\ \ \ }c@{\ \ \ }c@{\ \ \ }c@{\ \ \ }c@{\ \ \ }c@{\ \ \ }c@{\ }}
\hline
\hline	
Name & P $\pm$0.001d & M1/M2 & Comments \\
\hline	
LMC33-MHF87970 & 7.117 & 2.8 & SB2 \\
LMC33-MHF111340 & 1.074 & 1.2 & SB2 \\
LMC33-MHF127573 & 2.932 & $\simeq$1 & SB2, total eclipse\\
LMC33-MHF141891 & 2.975 & & SB1 \\
LMC33-MHF149652 & 1.458 & & SB1, total eclipse\\
\hline
SMC5\_977 & 3.128 & & SB2\\
SMC5\_4477 & 2.987 & & MOA P=4.482d\\
SMC5\_4534 & 4.051 & &  \\
SMC5\_13723 & 2.059 & & OGLE same P\\
SMC5\_20391 & 2.320 & & MOA same P\\
SMC5\_23571 & 3.534 & & Ellipsoidal bin.\\
SMC5\_23641 & 2.010 & & \\
SMC5\_24122 & 4.246 & & Eccentric bin.\\
SMC5\_49816 & 0.332 & & WUMa\\
SMC5\_74928 & 2.137 & & Ellipsoidal bin.\\
SMC5\_84353 & 1.557 & & \\
\textit{SMC5\_2807} & 454.959 & & \textit{Cool Sg. OGLE same P}\\
\textit{SMC5\_3789} & 2.087 & & \textit{Be bin.}\\
\textit{SMC5\_16461} & 54.317 & & \textit{Be Ellipsoidal bin.}\\
\hline
\end{tabular}
\label{binaires}
}
\end{table*}

\section{Photometric variability in Be stars}

As for galactic Be stars observed with Hipparcos (Hubert \& Floquet 1998)  and
for Be stars candidates in the SMC (Mennickent et al. 2002),  the short-term
variability  is often superimposed to a longer one which  can be complex: short-,
long-lived outbursts or fadings, periodic or  pseudo-periodic variations.
Therefore,  when the Be stars lightcurves are  not too complicated, 
we searched for  short-term variability  by using PDM and
FT+CLEAN algorithm methods.  Under these conditions we find that 13 Be stars 
present short-term  photometric variability (P$<$2.5d) with amplitude lower
than 0.1 magnitude. Results are given in Table~\ref{varBe}. 
The value of periods of short-term variations seem to be of the same order
in the MC and in our Galaxy (see Hubert \& Floquet 1998). We observed 6 stars
proposed by  Mennickent et al. (2002) as Be stars candidates and we confirm
their Be nature.  Due to their complex light curves it was not possible to
investigate  the short-term variability of Be stars in the sample
in the LMC.

\begin{table*}[tbph]
\centering
\scriptsize{
\caption{Characteristics of short-term photometric variability in Be stars 
in the Magellanic Clouds.
The first column gives the name of the star, the second column gives the 
period with an accuracy of $\pm$0.001d . The third column gives the
amplitude and the last column gives different comments about the Be star.}
\begin{tabular}{@{\ }c@{\ \ \ }c@{\ \ \ }c@{\ \ \ }c@{\ \ \ }c@{\ \ \ }c@{\ \ \ }c@{\ \ \ }c@{\ \ \ }c@{\ \ \ }c@{\ }}
\hline
\hline	
Name & P $\pm$0.001d & Amplitude & Comments \\
\hline	
SMC5\_3296 & 0.499 & 0.04 & \\
SMC5\_13978 & 1.685 & 0.04 & P$_{Bal}$=0.731d\\
SMC5\_14727 & 0.891 & 0.04 & \\
SMC5\_16523 & 1.547 & 0.09 & \\
SMC5\_16544 & 0.586 & 0.07 & \\
SMC5\_21152 & 1.015 & 0.04 & \\
SMC5\_37162 & 1.130 & 0.08 & \\
SMC5\_43413 & 1.000 & 0.06 & \\
SMC5\_82042 & 0.402 & 0.05 & \\
SMC5\_82941 & 1.600 & 0.1 & \\
MHF[S9]35238 & 0.754 & 0.07 & \\
MHF[S9]37842 & 0.846 & 0.05 & \\
MHF[S9]39981 & 0.783 & 0.04 & \\ 
\hline
\end{tabular}
\label{varBe}
}
\end{table*}

\end{document}